\documentclass[prc,nofootinbib,superscriptaddress]{revtex4-1}

\usepackage[utf8]{inputenc}
\usepackage[dvipsnames]{xcolor}
\usepackage{graphicx}
\graphicspath{{figures/}}

\usepackage{float}
\usepackage{bm}
\usepackage{amsmath, amsthm, amssymb, amsfonts}
\usepackage{bbm}
\usepackage{slashed}
\usepackage{physics}
\usepackage[colorlinks=true,allcolors = blue]{hyperref}
\usepackage{xspace}

\usepackage{ulem}
\usepackage{cancel}

\renewcommand{\emph}[1]{\textit{#1}}

\newcommand{\eftnopi}{EFT$_{\pi\hskip-0.40em /}$\xspace}
\newcommand{\nopicutoff}{\Lambda_{\not{\pi}}}
\newcommand{\mass}{M}
\newcommand{\Nc}{\ensuremath{N_c}\xspace}
\newcommand{\Qmu}{\widehat{\mu}}

\newcommand{\LB}{\left(}
\newcommand{\RB}{\right)}
\newcommand{\LSB}{\left[}
\newcommand{\RSB}{\right]}

\newcommand{\oneS}{{{}^{1}\!S_0}}
\newcommand{\threeS}{{{}^{3}\!S_1}}
\newcommand{\Ptwo}{{{}^{3}\!P_2}}
\newcommand{\threePone}{{{}^{3}\!P_1}}
\newcommand{\Pzero}{{{}^{3}\!P_0}}

\newcommand{\oneP}{{{}^{1}\!P_1}}

\newcommand{\threeD}{{{}^{3}\!D_1}}
\newcommand{\threeDtwo}{{{}^{3}\!D_2}}

\newcommand{\SP}{$S\text{-}P\!$\xspace}
\newcommand{\DP}{$P\text{-}D\!$\xspace}
\newcommand{\SD}{$S\text{-}D\!$\xspace}

\newcommand{\threeSonePL}{C^{(^3\!S^{[2]}_1-^1\!P^{[1]}_1)}}
\newcommand{\oneSthreePscalarL}{C^{(^1\!S^{[2]}_0-^3\!P^{[1]}_0)}_{\Delta I=0}}
\newcommand{\oneSthreePvectorL}{C^{(^1\!S^{[2]}_0-^3\!P^{[1]}_0)}_{\Delta I=1}}
\newcommand{\oneSthreePtensorL}{C^{(^1\!S^{[2]}_0-^3\!P^{[1]}_0)}_{\Delta I=2}}
\newcommand{\threeSthreePL}{C^{(^3\!S^{[2]}_1-^3\!P^{[1]}_1)}}

\newcommand{\threeSonePR}{C^{(^3\!S^{[0]}_1-^1\!P^{[3]}_1)}}
\newcommand{\oneSthreePscalarR}{C^{(^1\!S^{[0]}_0-^3\!P^{[3]}_0)}_{\Delta I=0}}
\newcommand{\oneSthreePvectorR}{C^{(^1\!S^{[0]}_0-^3\!P^{[3]}_0)}_{\Delta I=1}}
\newcommand{\oneSthreePtensorR}{C^{(^1\!S^{[0]}_0-^3\!P^{[3]}_0)}_{\Delta I=2}}
\newcommand{\threeSthreePR}{C^{(^3\!S^{[0]}_1-^3\!P^{[3]}_1)}}

\newcommand{\threeSoneP}{C^{(^3\!S_1-^1\!P_1)}_3}
\newcommand{\oneSthreePscalar}{C^{(^1\!S_0-^3\!P_0)}_{3,~\Delta I=0}}
\newcommand{\oneSthreePvector}{C^{(^1\!S_0-^3\!P_0)}_{3,~\Delta I=1}}
\newcommand{\oneSthreePtensor}{C^{(^1\!S_0-^3\!P_0)}_{3,~\Delta I=2}}
\newcommand{\threeSthreeP}{C^{(^3\!S_1-^3\!P_1)}_3}

\newcommand{\threeDoneP }{C^{(^1\!P_1-^3\!D_1)}_3}
\newcommand{\threeDthreePone}{C^{(^3\!P_1-^3\!D_1)}_3}
\newcommand{\oneDthreePscalar}{C^{(^3\!P_2-^1\!D_2)}_{3,~\Delta I=0}}
\newcommand{\oneDthreePvector}{C^{(^3\!P_2-^1\!D_2)}_{3,~\Delta I=1}}
\newcommand{\oneDthreePtensor}{C^{(^3\!P_2-^1\!D_2)}_{3,~\Delta I=2}}
\newcommand{\threeDthreePtwo}{C^{(^3\!P_2-^3\!D_2)}_3}

\newcommand{\twoway}{\overleftrightarrow{\nabla}}

\newcommand{\beq}{\begin{equation}}
\newcommand{\eeq}{\end{equation}}

\newcommand{\beqlist}{\begin{equation}\begin{aligned}}
\newcommand{\eeqlist}{\end{aligned}\end{equation}}

\newcommand{\CSP}{\ensuremath{C^{(SP)}}}
\newcommand{\CDP}{\ensuremath{C^{(PD)}_3}}
\newcommand{\CSD}{\ensuremath{C^{(\threeS-\threeD)}_2}}

\newcommand{\calA}{\mathcal{A}}
\newcommand{\calL}{\mathcal{L}}
\newcommand{\calO}{\mathcal{O}}

\newcommand{\hc}{\text{h.c.}}

\newcommand{\isospinvector}{\epsilon_{ab3}}
\newcommand{\isospintensor}{\mathcal{I}_{ab}}

\newcommand{\murange}{\ensuremath{|1/a - \mu| \ll \nopicutoff}\xspace}

\newcommand{\murangetriplet}{\ensuremath{|1/a^{(\threeS)} - \mu | \ll \nopicutoff}\xspace}

\newcommand{\NN}{$N \! N$\xspace}
\begin{document}
\title{Large-\Nc and renormalization group constraints on parity-violating low-energy coefficients for three-derivative operators in pionless effective field theory }
\author{Son T. Nguyen}
\email{stn7@phy.duke.edu}
\affiliation{Department of Physics, Duke University, Durham, NC 27708, USA}
\author{Matthias R.~Schindler}
\email{mschindl@mailbox.sc.edu}
\affiliation{Department of Physics and Astronomy, University of South Carolina, Columbia, SC 29208, USA}
\author{Roxanne P.~Springer}
\email{rps@phy.duke.edu}
\affiliation{Department of Physics, Duke University, Durham, NC 27708, USA}
\author{Jared Vanasse}
\email{jvanass3@fitchburgstate.edu}
\affiliation{Fitchburg State University, Fitchburg, MA 01420, USA}
\date{\today}
\begin{abstract}

We extend from operators  with one derivative to operators with three derivatives the analysis of two-body hadronic parity violation in a combined pionless effective field theory (\eftnopi) and large-\Nc expansion, where \Nc is the number of colors in quantum chromodynamics (QCD). In elastic scattering, these operators contribute to \SP and \DP wave transitions, with five operators and their accompanying low energy coefficients (LECs) characterizing the \SP transitions and six operators and LECs those in \DP transitions.
We show that the large-\Nc analysis  separates them into leading order in \Nc,  next-to-leading order in \Nc, etc.  Relationships among \eftnopi LECs emerge in the  large-\Nc expansion. We also discuss the renormalization scale dependence of these LECs. Our analysis can complement lattice QCD calculations and help prioritize future parity-violating experiments.

\end{abstract}
\maketitle
    
\section{Introduction}

Understanding weak interactions in few-nucleon systems may provide a window into how nonperturbative quantum chromodynamics (QCD) governs nuclear structure and behavior. Since parity is conserved in electromagnetic and strong interactions but not in weak interactions, parity-violating (PV) processes involving few nucleons can be used to isolate the  effects of the weak interactions, which are suppressed by about seven orders of magnitude relative to the strong interactions  (see, e.g., \cite{Zhu:2004vw,RamseyMusolf_looking_glass_2006, Haxton_HPV_2013,Schindler:2013yua,Gardner:2017xyl}  and their references for theoretical reviews and experimental status).
The nonperturbative nature of QCD at low energies makes direct calculations of hadronic PV observables challenging. While lattice QCD efforts in this direction  continue, they are still in their infancy (see, e.g., \cite{Gupta_latticeQCD,Beane_NP_latticeQCD, Lahde_Nuclear_lattice_EFT} for reviews). Effective field theories (EFTs) provide a model-independent approach for understanding  interactions among nucleons.  

EFTs  exploit any hierarchy of  scales  existing in a system. They retain dynamics that are relevant at the desired energy scale as well as underlying symmetries, while all high-energy/short-distance dynamics are encoded into  low-energy coefficients (LECs), which are not determined by these symmetries. The LECs may be  calculated from the underlying theory or fit to experimental data. At any given  order in the EFT expansion, only a finite number of LECs are relevant. Once these have been fixed by comparison with data, their values can be used to predict additional observables.

At momenta significantly below the pion mass ($ m_\pi \approx 140$ MeV), an EFT describing few-nucleon physics may be constructed in which  the only dynamical degrees of freedom are non-relativistic nucleons and possibly photons, neutrinos, etc.  In this pionless EFT (\eftnopi) \cite{Kaplan_KSW_1996,Kaplan_KSW_1998a, Kaplan_KSW_1998b, Chen:1999tn, Beane:crossing_border_2000, Bedaque:2002mn, vanKolck:2019vge}, the operators are organized in powers of $p/\nopicutoff$, where $p$ is a typical external momentum or momentum transfer in the system, and $\nopicutoff \sim m_\pi$ is the breakdown scale of \eftnopi. When restricted to two-nucleon processes, \eftnopi is written in terms of four-nucleon contact interactions, with terms involving an adequate number of derivatives to reach the desired level of precision. \eftnopi is well established for the low-energy regime of $E\lesssim 10$ MeV in the lab frame (or momentum transfers of $p \ll \nopicutoff$) (For examples, see Refs.~\cite{Kaplan_KSW_1996,Kaplan_KSW_1998a, Kaplan_KSW_1998b, Chen:1999tn, Beane:crossing_border_2000, Bedaque:2002mn, vanKolck:2019vge}).   

Another useful tool for understanding QCD is  the large-\Nc limit, where the number of colors \Nc is taken to be large (the physical value is $\Nc = 3$) \cite{HOOFT1974461, Witten:1979}. Combining the large-\Nc expansion with the \eftnopi expansion puts constraints on the LECs of \eftnopi.  
Given the success of this dual \eftnopi and large-\Nc expansion when used to analyze parity-conserving (PC) observables \cite{Kaplan_spin_flavor_largeN_1995, Kaplan_NN_in_1/N_1996, Schindler_2derivative_2018}, we have some confidence that it will be useful for analyzing PV processes as well.  
If the dual expansion can be used to indicate which PV terms are most important, it may help prioritize future PV experiments.  As an example of the application of this dual expansion, Ref.~\cite{Schindler_1derivative_2016} found that two isoscalar PV two-nucleon LECs  that are independent in the \eftnopi expansion are related in the large-\Nc expansion, with  $1/\Nc^2$ corrections. 
The approach of combining \eftnopi with large-\Nc has  been applied to PC nucleon-nucleon (\NN)  interactions~\cite{Kaplan_spin_flavor_largeN_1995, Kaplan_NN_in_1/N_1996, Schindler_2derivative_2018}, PV \NN  interactions~\cite{Schindler_1derivative_2016}, time-reversal violating \NN  interactions~\cite{Vanasse:2019fzl}, and interactions with external currents~\cite{Richardson:2020iqi}.

In this paper, we use the dual  \eftnopi and large-\Nc expansion to study the impact of three-derivative operators  on PV \NN interactions.
The leading (largest) PV \NN terms appear with one derivative and describe \SP wave transitions.
For elastic scattering, five independent three-derivative operators provide corrections to  these \SP wave transitions, and an additional six independent three-derivative operators contain the leading contributions to \DP wave transitions.   The dual expansion yields an estimate of which of these eleven operators will be dominant; 
the large-\Nc expansion suggests that only two of these \eftnopi LECs are both dominant and independent.  This allows us to obtain corrections to the leading \SP  LEC relationships found in Ref.~\cite{Schindler_1derivative_2016} and provides constraints for \DP LECs as well. However, at the moment there is no experimental evidence available to verify these predictions.

A subtlety of the large-\Nc analysis is that it applies not to observables but to LECs, which are typically subtraction-point ($\mu$) dependent. We therefore investigate the $\mu$ dependence of the LECs.  A renormalization group (RG) analysis shows that PV scattering does not require a new LEC until two orders past leading. Further, we see that the PV LEC RG equations allow general conclusions to be drawn about the behavior of higher order LECs, paralleling those found in the PC sector \cite{Kaplan_KSW_1998a,Kaplan_KSW_1998b,vanKolck:1998bw}.  Finally, for suitably small values of $\mu$ the RG results relate LECs in different channels.  However, previous results \cite{Kaplan_spin_flavor_largeN_1995,Schindler_2derivative_2018}  suggest that  at these small values of $\mu$ the large-\Nc constraints do not hold. 

At present, there exist two measurements of PV two-nucleon observables performed at the low energies where \eftnopi is valid. The longitudinal asymmetry in $\vec{p}p$ scattering was found 
in Refs.~\cite{Eversheim:1991tg, Haeberli:1995uz}. More recently, the NPDGamma Collaboration published the PV gamma-ray asymmetry arising from  polarized neutron capture on the proton~\cite{Blyth:2018aon}. Potential future experiments, e.g., at the high-intensity cold neutron beamline at the Spallation Neutron Source at the Oak Ridge National Laboratory  and the High-Intensity Gamma-ray Source  at the Triangle Universities Nuclear Laboratory \cite{Ahmed:2013jma,HIGS2} are expected to provide additional constraints on LECs. At the same time, there are attempts to calculate PV LECs using lattice QCD \cite{Beane_NP_latticeQCD,Wasem_latticeNPV,Tiburzi_isotensor_HPV}. The isotensor terms in particular present an opportunity for lattice QCD because the calculation does not involve disconnected (quark-loop) diagrams.
As shown in Refs.~\cite{Phillips_LargeN_NN_force_2015, Schindler_1derivative_2016},
 isotensor contributions are dominant in large-$\Nc$ counting; they are particularly attractive for lattice QCD and future experiments.

An outline of this paper is as follows: we begin with a brief review of the large-\Nc counting rules in Sec.~\ref{sec2}. In Sec.~\ref{sec3}, we introduce the PV Lagrangian expressed in two different bases; one in which the large-\Nc scaling is most transparent, the other describing the interactions in terms of partial-wave transitions.
We will refer to these as the large-\Nc and partial-wave bases, respectively. The large-\Nc scaling of the  partial-wave LECs is then provided. We  discuss the RG  behaviors of LECs and apply them to available experimental data  in Sec.~\ref{sec4}. Conclusions and an appendix showing an example Fierz procedure follow.


\section{Large-\Nc counting \label{sec2}}

Using the large-\Nc limit of QCD to understand hadron properties was first suggested by t'Hooft \cite{HOOFT1974461} and further developed by Witten \cite{Witten:1979}.  Assuming a sensible $\Nc \rightarrow \infty$ limit exists, $1/\Nc$ may serve as a useful expansion parameter for large, but finite \Nc.
Many calculations have demonstrated the capability of this method to make definite predictions of meson and baryon dynamics (see, e.g., \cite{Dashen:baryon_pion_coupling_1993,Jenkins_light_quark_1993,  Dashen:Nc_expansion_for_baryons_1994, Dashen:spin_flavor_structure_1995, Kaplan_spin_flavor_largeN_1995, Kaplan_NN_in_1/N_1996, Banerjee:2001js}).

In the large-\Nc approach, baryons consist of \Nc quarks. Quark confinement requires that the baryon wave function  be an SU$(\Nc)$ color singlet. Since the total wave function of the baryon must be completely antisymmetric, the baryon's ground state wave function is totally symmetric in spin and flavor components. This motivates the introduction of so-called ``bosonic quarks''  from which the color degrees of freedom are removed \cite{Dashen:baryon_pion_coupling_1993, Jenkins_light_quark_1993, Dashen:Nc_expansion_for_baryons_1994, Dashen:spin_flavor_structure_1995}. Neutrons and protons consist of valence up $(u)$ and down $(d)$ quarks, which can each be in two possible spin states $(\uparrow,\downarrow)$. Ref.~\cite{Dashen:spin_flavor_structure_1995}  showed that in the \Nc $\rightarrow \infty$ limit there is an SU(4) spin-flavor symmetry with $u\uparrow,u\downarrow,d\uparrow$, and $d\downarrow$ in the fundamental representation, where $u$ and $d$ are the bosonic quarks. 
In the large-\Nc limit, \NN interactions take the form of a Hartree Hamiltonian \cite{Witten:1979,Dashen:spin_flavor_structure_1995,Kaplan_NN_in_1/N_1996,Kaplan_spin_flavor_largeN_1995}:
\begin{equation}\label{Hartree}
    \Hat{H}=\Nc\sum_{n=0}^{\Nc}\sum_{s+t\leq n}v_{stn}(\vb{p},\vb{p}')\bigg(\frac{\hat{S}}{\Nc}\bigg)^s\bigg(\frac{\hat{I}}{\Nc}\bigg)^t\bigg(\frac{\hat{G}}{\Nc}\bigg)^{n-s-t},
\end{equation}
where $v_{stn}$ is a function of momenta  and scales at most as $O(\Nc^0)$ in the large-\Nc counting.
The operators in Eq.~\eqref{Hartree} and the identity are
\begin{equation}
\begin{aligned}
 \hat{S}_i =\hat{q}^\dagger\frac{\sigma_i\otimes \mathbbm{1}}{2}\hat{q}, ~~\hat{I}_a & = \hat{q}^\dagger\frac{\mathbbm{1}\otimes\tau_a}{2}\hat{q},~~
    \hat{G}_{ia} = \hat{q}^\dagger\frac{\sigma_i\otimes\tau_a}{4}\hat{q},~~
    \hat{\mathbbm{1}}=
    \hat{q}^\dagger \LB \mathbbm{1} \otimes \mathbbm{1}\RB \hat{q},
\end{aligned}
\label{eq:SIG}
\end{equation}
where $\hat{q}=(u,d)$ are bosonic quarks, and $\sigma_i$, $\tau_a$ are SU(2) Pauli matrices ($i,a=1,2,3$) acting on spin and isospin spaces, respectively.  
When evaluated between nucleon states, the large-\Nc counting rules for matrix elements of the spin-isospin  operators and the identity are  given by \cite{Dashen:spin_flavor_structure_1995, Kaplan_spin_flavor_largeN_1995,Kaplan_NN_in_1/N_1996}
\begin{equation}
    \begin{aligned}
        &\langle N'|\hat{S}_i/\Nc| N\rangle\sim\langle N'|\hat{I}_a/\Nc| N\rangle \lesssim \Nc^{-1}, \\
        &\langle N'|\hat{G}_{ia}/\Nc| N\rangle \sim \langle N|\hat{\mathbbm{1}}/\Nc| N\rangle  \lesssim \Nc^0,
    \end{aligned}
\label{eq5}
\end{equation}
where $N=(p,n)^T$ is the nucleon field.

The potential between two nucleons can be viewed as a matrix element of the above Hartree Hamiltonian \cite{Kaplan_NN_in_1/N_1996}. In the center-of-mass (c.o.m) frame 
\begin{equation}
    V=\langle \vb{p}'; -\vb{p}'|\hat{H}|\vb{p}; -\vb{p}\rangle.
\end{equation}
Two independent momentum variables can be defined
\cite{Kaplan_NN_in_1/N_1996},
\begin{equation}
    \vb{p}_+=\vb{p}+\vb{p}', \quad  \vb{p}_-=\vb{p}-\vb{p}'.
\end{equation}
How momenta scale with \Nc is a subject of much debate \cite{Witten:1979,Banerjee:2001js}. As discussed in Ref.~\cite{Banerjee:2001js}, one argument can be made using the meson-exchange potential derived from the Hartree Hamiltonian. Considering the $t$-channel, $\vb{p}_+$ only results from relativistic corrections and therefore contributes to the \Nc counting as $1/\mass \sim 1/\Nc$ since the nucleon mass $\mass$ scales as \Nc \cite{Witten:1979}. The large-\Nc counting of momenta in this channel then becomes
\begin{equation}
    \vb{p}_{-}\sim \Nc^0, \quad \vb{p}_{+}\sim \Nc^{-1}.  
    \label{eq6}
\end{equation}
The analysis in the $u$-channel is complementary, so we will restrict the discussion to the $t$-channel. Equations~\eqref{eq5} and \eqref{eq6}  are sufficient to systematically determine which spin-isospin structures occur at each large-\Nc order.
We will use this to describe two-nucleon PV scattering, retaining rotational and time-reversal invariance.
A concise overview of the transformation properties of various spin-isospin operators can be found in Tables I and II of Ref.~\cite{Samart_Time_resersal_violation_2016}. These constraints restrict the possible terms in the PV potential. In the next section, following the procedure used in Ref.~\cite{Schindler_1derivative_2016}, we present the \eftnopi PV Lagrangian with three-derivatives and obtain the large-\Nc scaling of the relevant LECs.

\section{\eftnopi Parity-violating Lagrangian with three-derivatives  \label{sec3}}

At three derivatives, the effective two-nucleon interactions are characterized by dimension-nine operators. The choice of basis used to express these operators is arbitrary. As long as the underlying symmetries are obeyed, different choices of bases must yield the same physics. However, as is often the case, some basis choices are more illuminating than others.

In this paper, we will consider the PV \eftnopi Lagrangian in two bases: The large-\Nc basis, expressed in the form $(N^\dagger {\cal O}_1 N)(N^\dagger {\cal O}_2 N)$; and the partial-wave basis, expressed in the form $(N^T {\cal O}_3 N)^\dagger(N^T {\cal O}_4 N)$, where the ${\cal O}_i$ are spin-isospin and momentum operators. While the physics might be more transparent in the partial-wave basis, the large-\Nc scaling can be most easily determined in the large-\Nc basis. 
Fierz identities can be used to transform between these two bases, but  this must be done with care in order to maintain the correct large-\Nc ordering of the operators.

\subsection{Large-\Nc basis Lagrangian}
The terms in the PV three-derivative Lagrangian that  appear at leading order (LO) in the large-\Nc expansion are given in the large-\Nc  basis by
\beqlist
    \calL_\text{LO}  =&~  C^{[3]}_{G.G} ~ \epsilon_{ijk} \nabla_i (N^\dagger \sigma_j\tau_a N)\nabla^2(N^\dagger \sigma_k\tau_a N)\\
    & + C'^{[3]}_{G.G} ~ \epsilon_{ijk} \mathcal{I}_{ab} \nabla_i (N^\dagger \sigma_j\tau_a N) \nabla^2 (N^\dagger \sigma_k\tau_b N),
    \label{LO_weinberg_Lag}
\eeqlist
corresponding to a potential 
\beqlist
    V_{\text{LO}} = & -iC^{[3]}_{G.G} \  \vb{p}_{-}^2 \ \vb{p}_{-}\cdot(\vec{\sigma}_1\times\vec{\sigma}_2)(\vec{\tau}_1\cdot\vec{\tau}_2)\\
    &- iC'^{[3]}_{G.G}  \ \vb{p}_{-}^2 \ \vb{p}_{-}\cdot(\vec{\sigma}_1\times\vec{\sigma}_2)(\tau_1^a\mathcal{I}_{ab}\tau_2^b).
    \label{eq:V-LO}
\eeqlist
Here, $\mathcal{I}_{ab} = \textnormal{diag}(1, 1, -2)$, and $C^{[3]}_{G.G}$, $C'^{[3]}_{G.G}$ are two LECs with subscripts chosen to reflect the operator structures they accompany. These two terms describe isospin transitions $\Delta I=0$ and $\Delta I=2$, respectively. The rules of Eqs.~(\ref{eq5}) and~\eqref{eq6} yield the \Nc scaling of the two couplings:
\begin{equation}
    C_{G.G}^{[3]}\sim \Nc, \quad C'^{[3]}_{G.G} \sim \Nc,
\end{equation}
where a factor of $\sin^2\theta_W\approx0.23$ \cite{Tanabashi:2018oca} in the isotensor coupling could be included because of matching at the weak scale \cite{Phillips_LargeN_NN_force_2015,Schindler_1derivative_2016}. However, it is not clear that this suppression survives the strong running to the low energy scales we consider. At next-to-leading order (NLO), scaling as $\Nc^0$ in large-\Nc, there are  four  $\Delta I =1$ operators,
\begin{widetext}
\beqlist
    \calL_{\text{NLO}}=&~\frac{1}{2}
    \widetilde{C}^{[3]}_{G.\sigma}~ \epsilon_{ijk}\LSB\nabla_i(N^\dagger\sigma_j\tau_3N)\nabla^2(N^\dagger\sigma_k N)+\nabla_i(N^\dagger\sigma_jN)\nabla^2(N^\dagger\sigma_k\tau_3 N)\RSB\\
    +&~ \frac{1}{2} \widetilde{C}^{[3]}_{G.\tau}~\epsilon_{ab3}\big[\nabla_i(N^\dagger\sigma_i\tau_aN)\nabla^2(N^\dagger\tau_bN)+\nabla_i(N^\dagger\tau_aN)\nabla^2(N^\dagger\sigma_i\tau_bN)\big]\\
    +&~ \widetilde{C}^{[3]}_{G.G}~ \epsilon_{ijk}\epsilon_{ab3}[(N^\dagger \sigma_i \tau_ai\twoway_j N)\nabla_k \nabla_l (N^\dagger \sigma_l \tau_b N)+(N^\dagger \sigma_l\tau_ai\twoway_j N)\nabla_k \nabla_l (N^\dagger \sigma_i \tau_b N)]\\
    +&~  \widetilde{C}^{[3]}_{G.1} ~ [(N^\dagger \sigma_i \tau_3 i\twoway_i N) \nabla^2 (N^\dagger N)-(N^\dagger i\twoway_i N)\nabla^2(N^\dagger \sigma_i \tau_3 N)].
    \label{NLO_weinberg_Lag}
\eeqlist
\end{widetext}
The tilde symbol indicates NLO in large-\Nc. The Galilean-invariant derivative $\twoway$ is defined by $N^\dagger \mathcal{O} \twoway_i N \equiv N^\dagger\mathcal{O}(\nabla_iN)-(\nabla_i N^\dagger)\mathcal{O} N$,  where $\calO$ is a spin-isospin operator. Again, the $\Delta I =1$ terms may be accompanied by a factor of $\sin^2 \theta_W$ \cite{Phillips_LargeN_NN_force_2015,Schindler_1derivative_2016} from matching at the weak scale.  The corresponding NLO potential is
\beqlist  
    V_\text{NLO} = & -\frac{i}{2}\widetilde{C}^{[3]}_{G.\sigma}~\vb{p}_{-}^2\, \vb{p}_{-}\cdot(\vec{\sigma}_1\times\vec{\sigma}_2)(\tau_1+\tau_2)^3 \\
    & - \frac{i}{2} \widetilde{C}^{[3]}_{G.\tau}~ \vb{p}_{-}^2\,\vb{p}_{-}\cdot(\vec{\sigma}_1 + \vec{\sigma}_2)(\vec{\tau}_1\times\vec{\tau}_2)^3\\
    & +\widetilde{C}^{[3]}_{G.G}~ \big[(\vb{p}_{+} \times \vb{p}_{-} )\cdot\vec{\sigma}_1\vb{p}_{-} \cdot\vec{\sigma}_2 \\
    & \qquad + \vb{p}_{-}\cdot\vec{\sigma}_1(\vb{p}_{+} \times \vb{p}_{-}) \cdot\vec{\sigma}_2\big] (\vec{\tau}_1\times\vec{\tau}_2)^3\\
    & + \widetilde{C}^{[3]}_{G.1}~\vb{p}_-^2\,\vb{p}_+\cdot(\vec{\sigma}_1\tau_1^3-\vec{\sigma}_2\tau_2^3).
    \label{eq:V-NLO}
\eeqlist

There are additional operators at higher orders in the large-\Nc expansion; however, in the following we restrict the discussion to the terms of Eqs.~\eqref{LO_weinberg_Lag} and \eqref{NLO_weinberg_Lag}, which are expected to be dominant in the large-\Nc expansion. The terms in Eqs.~\eqref{eq:V-LO} and \eqref{eq:V-NLO}, as well as the terms at next-to-next-to-leading order (NNLO) in the large-\Nc expansion, can be obtained from the expressions in Ref.~\cite{Phillips_LargeN_NN_force_2015} by expanding their functions $U_P^i(\vb{p}_{-})$ and $U_D^1(\vb{p}_{-})$ in $\vb{p}_{-}$ and retaining all terms with three powers of momentum.

\subsection{Partial-wave basis Lagrangian}

The PV \NN interactions can also be expressed in terms of the mixing of odd and even partial waves characterized by the spectroscopic notation $^{2S+1}\!L_J$. During the scattering process, the total angular momentum (orbital plus spin angular momentum) quantum number $J=0,1, 2$ of the two-nucleon system must be conserved. The three-derivative PV \NN interactions can be written in the partial-wave basis as
\begin{widetext}
\begin{align}
    \calL^{[3]}_{SP}  & =  \frac{1}{8}\Big[\threeSonePL (N^TP_i\twoway^2 N)^\dagger(N^TP_0 i\twoway_iN) \nonumber \\
    & \qquad  +\oneSthreePscalarL (N^T P_a \twoway^2 N)^\dagger (N^T P_{i,a} i\twoway_i N) \nonumber \\
    & \qquad  +\oneSthreePvectorL \isospinvector (N^T P_a \twoway^2 N)^\dagger (N^T P_{i,b} \twoway_i N) \nonumber \\
    & \qquad  +\oneSthreePtensorL \isospintensor (N^T P_a \twoway^2 N)^\dagger (N^T P_{i,b} i\twoway_i N) \nonumber \\
    & \qquad + \threeSthreePL \epsilon_{ijk} (N^T P_i\twoway^2 N)^\dagger (N^T P_{k,3} \twoway_j N)  \nonumber \\
    & \qquad  +\threeSonePR (N^TP_iN)^\dagger (N^T P_0 i\twoway_i \twoway^2N) \nonumber  \\
    & \qquad  +\oneSthreePscalarR (N^T P_a N)^\dagger (N^T P_{i,a} i\twoway_i \twoway^2 N) \nonumber \\
    & \qquad  +\oneSthreePvectorR \isospinvector (N^T P_a N)^\dagger (N^T P_{i,b} \twoway_i \twoway^2 N) \nonumber \\
    & \qquad  +\oneSthreePtensorR \isospintensor (N^T P_a N)^\dagger (N^T P_{i,b} i\twoway_i \twoway^2 N) \nonumber \\ 
    & \qquad  +\threeSthreePR \epsilon_{ijk} (N^TP_iN)^\dagger (N^T P_{k,3} \twoway_j \twoway^2 N) \Big] + \mathrm{h.c.}, \label{eq:3derLSP}  \\[2mm]
    \calL^{[3]}_{PD}  & = \frac{1}{8}\bigg[\threeDoneP(N^T P_i \twoway_x\twoway_y N)^\dagger(N^TP_0i\twoway_j N) \LB \delta_{ix}\delta_{jy} - \frac{1}{3}\delta_{ij}\delta_{xy} \RB \nonumber  \\
    & \qquad  +\threeDthreePone \epsilon_{ijk} (N^T P_l \twoway_x\twoway_y N)^\dagger (N^T P_{k,3}\twoway_jN) \LB \delta_{lx} \delta_{iy} - \frac{1}{3} \delta_{il} \delta_{xy} \RB \nonumber  \\
    & \qquad  +\oneDthreePscalar (N^T P_a\twoway_x\twoway_y N)^\dagger (N^T P_{i,a} i\twoway_j N)\LB\delta_{ix}\delta_{jy}  - \frac{1}{3}\delta_{ij}\delta_{xy} \RB \nonumber  \\
    & \qquad  +\oneDthreePvector \isospinvector (N^TP_a\twoway_x\twoway_yN)^\dagger (N^TP_{i,b}\twoway_j N) \LB \delta_{ix}\delta_{jy}  - \frac{1}{3}\delta_{ij}\delta_{xy} \RB \nonumber \\
    & \qquad  +\oneDthreePtensor \isospintensor (N^TP_a\twoway_x\twoway_yN)^\dagger (N^TP_{i,b} i\twoway_j N) \LB \delta_{ix}\delta_{jy}  - \frac{1}{3}\delta_{ij}\delta_{xy} \RB \nonumber \\
    & \qquad  +\threeDthreePtwo \epsilon_{ijk} (N^T P_j\twoway_k\twoway_l N)^\dagger (N^T P_{m,3}\twoway_n N)(\delta_{im}\delta_{ln} + \delta_{in}\delta_{lm}) \bigg] + \mathrm{h.c.}
    \label{eq:3derLPD} 
\end{align}
\end{widetext}
On the left hand side, the subscript denotes  the partial waves involved in the transition, and the superscript indicates the number of derivatives in the operators. On the right hand side, the LEC labels are chosen to echo the incoming/outgoing partial waves, and the superscript $[k]$ on a particular channel indicates the number of derivatives acting on that partial-wave channel.  $\Delta I$ denotes the isospin structure of the operator. The operators between nucleon fields are projectors onto the required combination of spin and isospin. The partial-wave projectors and momentum structures are provided in Ref.~\cite{Fleming_PW_projectors_1999},
\beqlist
    &P_i  = \frac{1}{\sqrt{8}} \sigma_2 \sigma_i \tau_2, &
    &P_a = \frac{1}{\sqrt{8}}\sigma_2\tau_2\tau_a,\\
    &P_{i,a}  =  \frac{1}{\sqrt{8}}\sigma_2\sigma_i\tau_2\tau_a, & &P_0 = \frac{1}{\sqrt{8}}\sigma_2\tau_2.
    \label{proj}
\eeqlist

In general, the magnitudes of the relative momenta of the incoming and outgoing nucleons are not equal, i.e., $|\vb{p}| \neq |\vb{p}'|$. But for elastic scattering, the two operators with different placements of $\twoway^2$ in each \SP transition cannot be distinguished; the LECs occur in the linear combination $C^{(S^{[2]}-P^{[1]})}$ + $C^{(S^{[0]}-P^{[3]})}$. Going forward, we will use the new set of LECs
\beq
C_3^{(SP)}\equiv C^{(S^{[2]}-P^{[1]})}+C^{(S^{[0]}-P^{[3]})}.
\eeq
A similar issue has been seen in the discussion of the two four-derivative $S$-wave operators $C_4$ and $\widetilde{C}_4$ in the PC sector \cite{RUPAK_bigbang_2000,Chen:1999tn}.

\subsection{Large-\Nc counting of partial-wave LECs \label{subsecIIIC}}
It is unclear how to directly count the large-\Nc behavior of the partial-wave basis terms appearing in  Eqs.~\eqref{eq:3derLSP} and \eqref{eq:3derLPD}; instead, the counting rules discussed in Sec.~\ref{sec2} are applied in the large-\Nc basis, and Fierz transformations are used to map the scaling to the partial-wave basis operators (see Appendix \ref{app2} for details). The large-\Nc scaling of the LECs, as well as the mapping between the LECs of the large-\Nc basis and the partial-wave basis, are shown in Table \ref{table1}.\footnote{References \cite{Phillips_LargeN_NN_force_2015,Schindler_1derivative_2016} include factors of $\sin^2 \theta_W$ in the counting of isovector and isotensor terms. Since these factors originate from matching at the weak scale and may be significantly modified by the nonperturbative running to the scales of \eftnopi, we do not show them in this table.}

\setlength{\tabcolsep}{6pt}
\renewcommand{\arraystretch}{1.8}
\begin{table}[!ht]
    \centering
    \begin{tabular}{c c c c c}\hline\hline
    LEC& & LO in \Nc, $O$(\Nc) & & NLO in \Nc, $O(\Nc^0)$ \\\hline
    $C^{(^3S_1-\oneP)}_3$ & = & $32C^{[3]}_{G.G}$  & + & $0$ \\
    $C^{(\oneS-\Pzero)}_{3,~\Delta I=0}$ & = & $\frac{32}{3}C^{[3]}_{G.G}$ &+ &$0$ \\[1mm]
    $C^{(\oneS-\Pzero)}_{3,~\Delta I=1}$ & = & $0$ & + & $\frac{16}{3}(2\widetilde{C}^{[3]}_{G.\sigma}  +\widetilde{C}^{[3]}_{G.1})$ \\[1mm]
    $C^{(\oneS-\Pzero)}_{3,~\Delta I=2}$ & = & $-\frac{64}{3}C'^{[3]}_{G.G}$  & + & $0$\\[1mm]
    $C^{(^3S_1-\threePone)}_3 $ & = & $0$ & + & $-\frac{16}{3} (2\widetilde{C}^{[3]}_{G.\tau} + \widetilde{C}^{[3]}_{G.1})$ \\\hline
    $\threeDoneP$ & = & $24 C^{[3]}_{G.G}$  & + & $0$ \\
    $\threeDthreePone$ & = &  $0$ & + &
    $4(\widetilde{C}_{G,\tau}^{[3]}-\widetilde{C}_{G,1}^{[3]}-6\widetilde{C}_{G,G}^{[3]})$ \\
    $\oneDthreePscalar$ & = & $8 C^{[3]}_{G.G}$  & + & $0$ \\[1mm]
    $\oneDthreePvector$ & = & $0$ & + & $  8(\widetilde{C}^{[3]}_{G.\sigma}  -\widetilde{C}^{[3]}_{G.1}) $    \\[1mm]
    $\oneDthreePtensor$ & = & $-16C'^{[3]}_{G.G}$ & + & $0$ \\[1mm]
    $\threeDthreePtwo$ & = & $0$ & + & $4(-\widetilde{C}_{G,\tau}^{[3]}+ \widetilde{C}_{G,1}^{[3]} - 2\widetilde{C}_{G,G}^{[3]})$ \\ \hline\hline
    \end{tabular}
    \caption{Large-\Nc scaling of and mapping betweIoe large-\Nc and partial-wave bases.}
    \label{table1}
\end{table}


The ratios of the \eftnopi \SP LECs in the three-derivative sector that are predicted at LO in the large-\Nc expansion agree with the ratios found in the one-derivative sector in Ref.~\cite{Schindler_1derivative_2016}. In particular, the two isoscalar LECs are not independent at LO in the  large-\Nc expansion. Their ratio is given by
\beq
    \frac{C^{(\threeS-\oneP)}_3}{C^{(\oneS-\Pzero)}_{3, ~\Delta I=0}} = 3\LSB 1 + O\LB \frac{1}{\Nc^2}\RB \RSB.
\end{equation}
Corrections to this identity are $O(1/\Nc^2)$ because all operators suppressed by a single factor of $1/\Nc$ are isovector.

For the six \DP wave transitions, the large-\Nc counting rules predict that $\threeDoneP$, $ \oneDthreePscalar$, and $\oneDthreePtensor$ are dominant.  We see that the large-\Nc behavior of the \DP LECs follows the pattern established by the \SP LECs: the $\Delta I=0$ and $\Delta I=2$ LECs are $O(\Nc)$, and only one $\Delta I=0$ LEC is independent. The relationship among the $\Delta I=0$ LECs in the large-\Nc expansion is predicted to be
\beq
    \frac{\threeDoneP}{\oneDthreePscalar} = 3\LSB 1 + O\LB \frac{1}{\Nc^2}\RB \RSB.
\end{equation}

At LO in the large-\Nc expansion, the results of Table \ref{table1} also predict relationships between \DP LECs and the three-derivative \SP LECs, such as
\beq
    \frac{\threeDoneP}{\threeSoneP} = \frac{3}{4}\LSB 1 + O\LB \frac{1}{\Nc^2}\RB \RSB,
\eeq
and
\beq
    \frac{\oneDthreePscalar + \oneDthreePvector + \oneDthreePtensor}{\oneSthreePscalar + \oneSthreePvector + \oneSthreePtensor } = \frac{3}{4}\LSB 1 + O\LB \frac{1}{\Nc}\RB \RSB.
\label{Eq:SPDrel}
\eeq
Consistent with what is observed in Refs.~\cite{Dashen:Nc_expansion_for_baryons_1994,Dashen:spin_flavor_structure_1995, Manohar:1998xv, Phillips_LargeN_NN_force_2015}, we find that the correction within a given channel is  $1/\Nc^2$. The correction in Eq.~\eqref{Eq:SPDrel} is  $O(1/\Nc)$ since the expression  contains isovector LECs in addition to isoscalar and isotensor LECs.

In addition, at NLO, we obtain a large-\Nc prediction for a ratio in the $\Delta I =1$ sector between \SP and \DP channels, 
\beq
\frac{2\oneDthreePvector -\threeDthreePone + 3\threeDthreePtwo}{\oneSthreePvector + \threeSthreeP} =\frac{3}{2}\LSB 1 + O\LB \frac{1}{\Nc^2}\RB \RSB.
\label{eq:SPD-I1}
\eeq
While we have not considered terms beyond NLO in the large-\Nc expansion here, as shown in Ref.~\cite{Phillips_LargeN_NN_force_2015} all terms at NNLO in large-\Nc are either isoscalar or isotensor; thus corrections to the isovector identity of Eq.~\eqref{eq:SPD-I1} are suppressed by $1/\Nc^2$.

The impact of this section is that while there are eleven independent operators with three-derivatives in \eftnopi, when the accompanying LECs are ranked by powers of 1/\Nc, the eleven LECs reduce to just two that are leading in both \eftnopi and a large-\Nc expansion.  If nature agrees, this could provide a powerful reduction in the number of experiments needed to understand PV in \NN scattering at every order in this dual expension.

However, LECs are not observables and are generally functions of the renormalization parameter $\mu$. Because the large-\Nc predictions considered in this paper apply to LECs and not observables, care must be taken to choose a $\mu$ that both illuminates the large-\Nc behavior and also obeys the power counting principles used in \eftnopi.  This will be discussed in the next section.   

\section{Renormalization group behavior of PV lecs \label{sec4}}

\subsection{Reviewing the parity-conserving renormalization group behavior}

While observables must be independent of renormalization scales and schemes, the LECs of a theory in general depend upon a renormalization parameter. In the power divergence subtraction (PDS) scheme \cite{Kaplan_KSW_1998a,Kaplan_KSW_1998b} used here, this parameter is the subtraction point $\mu$. The dependence of the LECs on $\mu$ can be determined by requiring that observables are $\mu$ independent. In the PC sector, this $\mu$ dependence has been analyzed in detail in, for example, Refs.~\cite{Kaplan_KSW_1998a, Kaplan_KSW_1998b,Birse:_RG_treatment_1998dk,Mehen:Momentum_subtraction_scheme_1999,Chen:1999tn, vanKolck:1998bw, RUPAK_bigbang_2000}.
The PC Lagrangian, including the $\oneS$, $\threeS$, and $\threeS-\threeD$ channels up to four derivatives, is (see,  e.g., Ref.~\cite{Chen:1999tn})\footnote{Our LEC $C_4^{\prime(\threeS)}$ is labeled $^{\slash \! \! \! \pi}C_4^{(\threeS)}$ in Ref.~\cite{Chen:1999tn}. Otherwise the notation is the same except for the absence of the $\slash \! \! \! \pi$ designation.}
\beqlist
    \calL^{PC} = & - C^{(\threeS)}_0(N^TP_i N)^\dagger(N^TP_i N)  - C^{(\oneS)}_0 (N^TP_a N)^\dagger (N^TP_a N)\\
    & + \frac{1}{8} C^{(\threeS)}_2 \LSB (N^TP_iN)^\dagger(N^T P_i\twoway^2 N) + \mathrm{h.c.} \RSB\\
    & + \frac{1}{8}C^{(\oneS)}_2 \LSB (N^TP_aN)^\dagger(N^TP_a\twoway^2 N) + \mathrm{h.c.} \RSB \\
    & + \frac{1}{4} \CSD \LSB(N^TP_iN)^\dagger (N^TP_j \twoway_x \twoway_y N) \LB \delta_{ix}\delta_{jy} -\frac{1}{3}\delta_{ij}\delta_{xy} \RB + \mathrm{h.c.}\RSB \\
    & - \frac{1}{16} C_4^{\prime(\threeS)}  (N^T P_i \twoway^2 N)^\dagger (N^T P_i \twoway^2 N) -\frac{1}{32} \widetilde{C}_4^{(\threeS)} \LSB (N^TP_iN)^\dagger(N^T P_i\twoway^4 N) + \mathrm{h.c.} \RSB \\
    & - \frac{1}{16} C_4^{\prime(\oneS)}  (N^T P_a \twoway^2 N)^\dagger (N^T P_a \twoway^2 N) -\frac{1}{32} \widetilde{C}_4^{(\oneS)} \LSB (N^TP_aN)^\dagger(N^T P_a\twoway^4 N) + \mathrm{h.c.} \RSB.
\label{eq24}
\eeqlist
The LECs may be determined by matching the \NN scattering amplitude to the effective range expansion about $p=0$, yielding \cite{Kaplan_KSW_1998a,Kaplan_KSW_1998b}
\begin{align}
    C_0(\mu) &= \frac{4\pi}{M}\bigg(\frac{1}{-\mu +1/a}\bigg),\label{C0scaling}\\
    C_2(\mu) &= \frac{4\pi}{M}\bigg(\frac{1}{-\mu +1/a}\bigg)^2  \ \frac{r_0}{2},
    \label{C2scaling}
\end{align}
where $a$ is the scattering length and $r_0$ is the effective range for the $\oneS$ or $\threeS$ channels, as needed. In the following we will continue to suppress the $\threeS/\oneS$ label on quantities such as $C_0$, $C_2$, $a$, and $r_0$ if they are unnecessary.
The running of the {\SD}-mixing LEC $\CSD$ is governed by the RG equation \cite{Kaplan_KSW_1998a, Fleming_SD_Mixing_2000}

\beq
\mu\dv{\mu}\CSD(\mu) = \frac{\mu M}{4\pi}C_0^{(\threeS)}(\mu)\CSD(\mu).
\label{eq:CSD_beta_func}
\eeq
Using $E_1^{(2)}= \eta_{SD}  \,(a^{(\threeS)})^2 + \cdots $ \cite{CHEN_Suppressed_amplitudes_1991}, where $\eta_{SD}$ is the asymptotic $SD$ mixing ratio,
 as a boundary condition for the solution to  Eq. \eqref{eq:CSD_beta_func}, $\CSD$ is given by~\cite{Chen:1999tn,Vanasse:2013sda} 
\begin{equation}
    \CSD(\mu)=-E_1^{(2)}\frac{6\sqrt{2}\pi}{M(\mu-\gamma)}=E_1^{(2)}\frac{3}{\sqrt{2}}C_0^{(\threeS)}(\mu)\ .
\end{equation}

\begin{figure}[h!]
  \centering
  \includegraphics[width=0.7\textwidth]{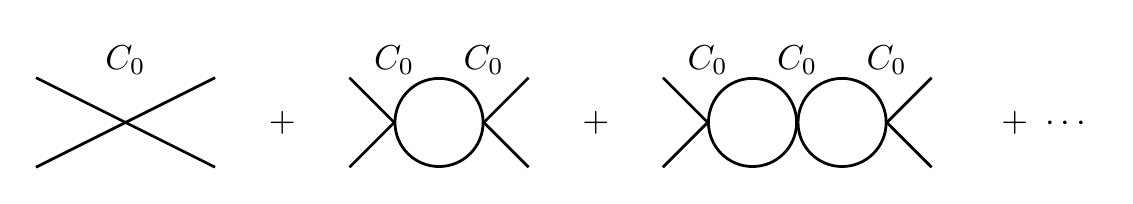}
  \caption{Diagrams contributing to PC \NN interactions at LO. $C_0$ can be the coupling in either the $\threeS$ or $\oneS$ channel.}
  \label{fig:bubble_sum}
\end{figure}

\eftnopi power counting is based on a small expansion parameter $Q/\nopicutoff$, where $\nopicutoff \sim m_\pi$ is the breakdown scale of \eftnopi and $Q\ll \nopicutoff$. Taking external momenta $p\sim Q$, the summation of the infinite series of loop diagrams in Fig.~\ref{fig:bubble_sum} can be justified in the PDS scheme if \murange.
While most applications of the \eftnopi power counting assume that \murange,  Refs.~\cite{Kaplan_spin_flavor_largeN_1995,Mehen:1999qs,CalleCordon:2008cz,CalleCordon:2009ps,RuizArriola:2016vap,Schindler_2derivative_2018} observe that the large-\Nc predictions do not agree with experiment unless $\mu \gtrsim  m_\pi$. A power counting for these larger values of $\mu$, including the justification for resumming the diagrams in Fig.~\ref{fig:bubble_sum}, can be maintained \cite{EPELBAUM:Wilson_RG_2017}; the corresponding LECs are of natural size, but the ordering of the perturbative expansion is modified compared to the case \murange.
In this section we will assume that we are in the region \murange, with $\mu \sim Q$ as well as $|1/a-\mu| \sim Q$. This leads to significant simplifications for certain PV LECs. However, the LEC relationships obtained in the present section are likely not  compatible with the LEC relationships obtained in the previous section on large-\Nc because of the different sizes of $\mu$ required.  In what follows we will keep the $\mu$ dependence explicit to emphasize this point. 
Because the power counting for \eftnopi deviates from naive dimensional analysis in the $S$-waves, it is important to keep distinct the mass dimension versus the power counting of operators in \eftnopi.
For example, both $C_2$ and $\CSD$ have mass dimension $-4$, but their dependence on dimensionful scales is different.
From Eq.~\eqref{C2scaling} and assuming that the effective range $r_0 \sim 1/\nopicutoff$, $C_2(\mu) \sim \frac{1}{M \Qmu^2 \nopicutoff}$, where 
\begin{equation}
\Qmu = \frac{1}{a}-\mu \ .
\end{equation}
On the other hand, $\CSD \sim \frac{1}{M \Qmu \nopicutoff^2} \ . $

At higher orders in the \eftnopi expansion, additional LECs contribute to $S$-wave scattering, which can be matched to higher orders in the ERE.
For elastic scattering, the two LECs $C^\prime_4$ and $\widetilde{C}_4$ only contribute in the linear combination
\beqlist
 C_4(\mu) & = C^\prime_4(\mu) + \widetilde{C}_4(\mu) \\
& = \frac{4 \pi}{M} \frac{r_0^2}{4}\left(
\frac{1}{-\mu+ 1/a} \right)^3 + \frac{4 \pi}{M}\frac{r_1}{2\nopicutoff^2 } \left(\frac{1}{ -\mu+ 1/a}  \right)^2 \ \ ,
\label{eq:C4}
\eeqlist
where $r_1/\nopicutoff^2$ is the shape parameter in the relevant (spin-singlet or spin-triplet) channel \cite{Kaplan_KSW_1998a}.
We are again suppressing $^1S_0$ and $^3S_1$ labels as Eq.~\eqref{eq:C4} is valid for both. If both $r_0$ and $r_1$ are of natural size, that is, if they both scale as $1/\nopicutoff$,  the first term in Eq.~\eqref{eq:C4} scales as 
$\frac{1}{M \Qmu^3 \nopicutoff^2}$ and the second as $\frac{1}{M \Qmu^2 \nopicutoff^3}$ , where again $\Qmu=1/a-\mu$. So long as \murange, the second term is suppressed relative to the first by a factor of $\Qmu/\nopicutoff$.
If this term is neglected, the LEC $C_4$ is entirely determined from the scattering length and the effective range and no new parameter enters.
In general,  if  \murange  and all parameters aside from the scattering length are of natural size (that is, they scale with inverse mass dimension $\nopicutoff$), the ``leading" (in a $\Qmu/\nopicutoff$ expansion) behavior for $C_{2n}$ is $\frac{1}{M \Qmu^{n+1} \nopicutoff^n }$, where $n>0$ and $2n$ is the number of derivatives associated with the operator for the $C_{2n}$ LEC \cite{Kaplan_KSW_1998a, Bedaque:2002mn}.

\subsection{\SP transitions at one and three derivatives}

Analogous arguments can be applied in the PV sector. The relevant one-derivative terms in the PV Lagrangian are \cite{Phillips:PV_EFT_2008, Schindler_1derivative_2016}
\beqlist
    \calL^{[1]}_{SP}=-\frac{1}{2} & \Big[ C^{(\threeS-\oneP)}_1(N^TP_iN)^\dagger(N^TP_0i\twoway_iN)\\
    &+ C^{(\oneS-\Pzero)}_{1,~\Delta I=0} (N^TP_aN)^\dagger(N^T P_{i,a} i\twoway_iN)\\
    &+ C^{(\oneS-\Pzero)}_{1,~\Delta I=1} \isospinvector (N^TP_aN)^\dagger(N^TP_{i,b}\twoway_i N)\\
    &+ C^{(\oneS-\Pzero)}_{1,~\Delta I=2} \isospintensor (N^TP_aN)^\dagger(N^TP_{i,b}i\twoway_iN)\\
    &+ C^{(\threeS-\threePone)}_1\epsilon_{ijk}(N^TP_iN)^\dagger(N^T P_{k,3}\twoway_j N)\Big] + \hc,
    \label{eq:onederivative}
\eeqlist
where the subscript ``1" on the LECs indicates the number of derivatives in the corresponding operator. The one-derivative Lagrangian in Eq.~\eqref{eq:onederivative} is different from the one used in Refs.~\cite{Phillips:PV_EFT_2008, Schindler_1derivative_2016} by an overall normalization factor of 16. The reason for this is to ensure a  normalization of the operators analogous to the three-derivative operators used in our large-$\Nc$ analysis.
\begin{figure}[ht!]
    \centering
    \includegraphics[width=0.7\linewidth]{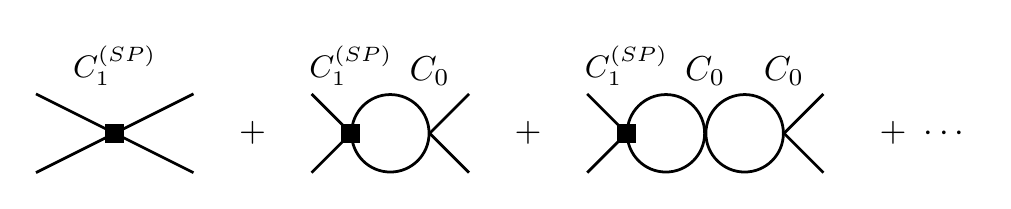}
    \caption{The LO PV \NN scattering diagrams. $\CSP_1$ is the leading weak interaction LEC and the $C_0$ LECs are PC. $\CSP_1$ could be any of the five   LECs from Eq.~(\ref{eq:onederivative}) and $C_0$ is $\oneS$ or $\threeS$ as needed.} \label{fig4}
\end{figure}
All diagrams in Fig.~\ref{fig4} are of the same order and must be summed to reproduce the LO PV scattering amplitude. This sum is a geometric series and simplifies to
\beqlist
    \calA^{[1]}_{SP} \propto &-p\CSP_1\frac{1}{1-I_0C_0} \ ,
    \label{eq:leading_PV_amplitude}
\eeqlist
where in the PDS subtraction scheme the loop integral $I_0$ is given by \cite{Kaplan_KSW_1998a}
\beqlist
    I_0 = &-i \LB\frac{\mu}{2}\RB^{4-D}\int\frac{d^Dq}{(2\pi)^D}\frac{i}{\frac{E}{2}+q_0-\frac{{\bf q}^2}{2M}+i\epsilon}\frac{i}{\frac{E}{2}-q_0-\frac{{\bf q}^2}{2M}+i\epsilon}\\
    \stackrel{\text{PDS}}{=} & -\frac{M}{4\pi}(\mu+ip), 
\eeqlist
with $E$ the total energy,  $p=\sqrt{ME}$ the relative momentum of the nucleons in the c.o.m.~frame, and $D$  the dimensions of spacetime used in dimensional regularization.
The scattering amplitude for each channel also contains the operator structure and an overall factor, which cancels out in the RG analysis. As discussed in Ref.~\cite{Phillips:PV_EFT_2008}, the scattering amplitude must be independent of $\mu$, which requires that
\begin{equation}
    \mu\dv{\mu} \frac{\CSP_1}{C_0}=0,
    \label{eq:RG_PVLO}
\end{equation}
or that the ratio $\CSP_1/C_0$ is $\mu$-independent. Thus the RG running of the  single-derivative PV LEC is entirely determined by that of the LO PC LEC.\footnote{At $\mu= 0$, $C_0(0) = 4\pi a/M$ from Eq.~\eqref{C0scaling}. This implies that at very low energies, the PV LEC becomes proportional to the scattering length, reminiscent of the Danilov parameters \cite{Danilov:1965,Danilov:1971fh}.}
The solution of Eq.~\eqref{eq:RG_PVLO} requires an undetermined integration constant that encodes the short-distance PV physics for each LEC.  Assuming that only the $S$-wave scattering lengths are unnatural, this implies that 
\beq
C_1^{(SP)}\sim\frac{1}{\nopicutoff}C_0\sim\frac{1}{M \Qmu \nopicutoff} \ .
\label{eq:CSP_scaling}
\eeq
However, while Eq.~\eqref{eq:CSP_scaling} provides information on the scaling of this PV LEC with nucleon mass $M$, subtraction point $\mu$, and cutoff scale $\nopicutoff$, there is always a dimensionless number that remains unspecified.  In the present case, where we are showing the relationship between the scaling of a PV LEC and the scaling of a PC LEC, this dimensionless number involves $G_F m_\pi^2 \approx 10^{-7}$, where $G_F$ is Fermi's coupling constant. This reflects the very different size of PV versus PC interactions, despite the fact that they scale the same with respect to $\mu$.
 
\begin{figure}[h]
    \centering
    \includegraphics[width=0.7\linewidth]{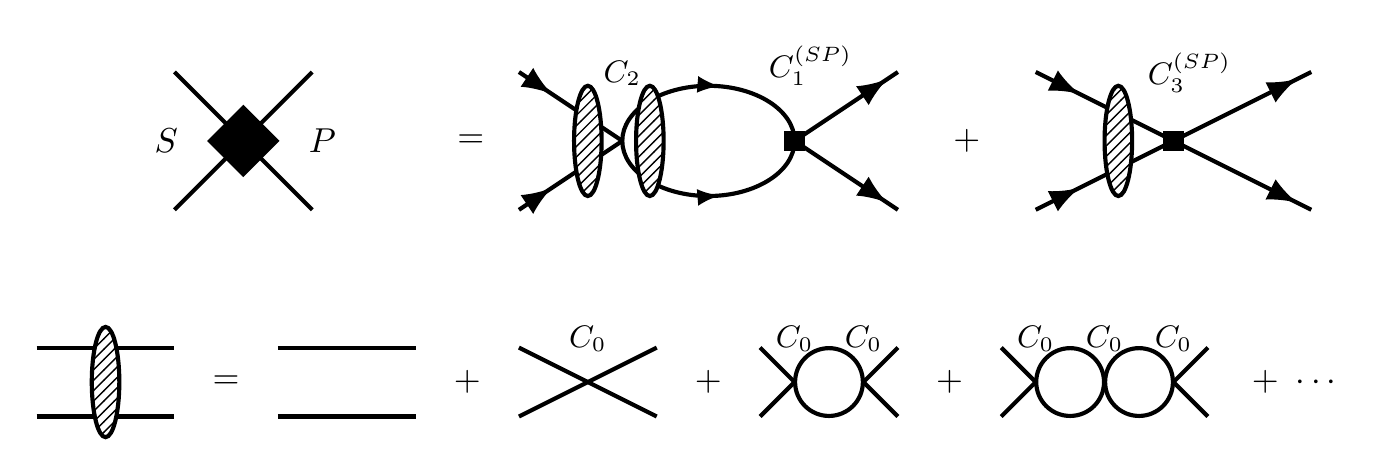}
    \caption{ The  diagrams contributing to the three-derivative corrections to \SP wave transitions in \eftnopi. The small squares denote insertions of PV operators with couplings $\CSP_1$ or $\CSP_3$ as indicated. The hashed ovals represent the strong $S$-wave re-scattering, which is the sum of a non-interaction and the LO PC \NN interactions. }
   \label{three_derivative_diagrams_SP}
\end{figure}

The contributions to elastic PV \SP wave scattering from the three-derivative \eftnopi terms of  Eq.~\eqref{eq:3derLSP} are shown in Fig. \ref{three_derivative_diagrams_SP}.

The \SP  scattering amplitude from three-derivative operators is 
\beq
    \calA_{SP}^{[3]} \propto - p^3 \CSP_1 \frac{I_0C_2}{(1-I_0C_0)^2} - p^3C_3^{(SP)} \frac{1}{1-I_0C_0}  \ .
    \label{AmpSP3}
\eeq
Each term in Eq.~\eqref{AmpSP3} individually depends on $\mu$. Requiring that the derivative of $\calA_{SP}^{[3]}$ with respect to $\mu$ vanishes yields
\beq
\mu\dv{\mu} \CSP_3(\mu) = \frac{\mu M}{4\pi}\LSB \CSP_1(\mu) \ C_{2}(\mu) + \CSP_3(\mu) \ C_{0}(\mu) \RSB.
\label{eq:C3SPmudep2}
\eeq
The solution to this equation is 
\beq
C_3^{(SP)}(\mu) = \frac{\CSP_1(\mu)}{C_0(\mu)} C_{2}(\mu) + \bar{C}_3^{(SP)}\ C_0(\mu),
\label{CSP3}
\eeq
where $\bar{C}_3^{(SP)}$ is a new $\mu$-independent integration constant that needs to be fixed from a lattice QCD calculation or from comparison to data. Currently insufficient data is available to do so. (The general structure of $C_3^{(SP)}(\mu)$ in Eq.~\eqref{CSP3} was already anticipated in Ref.~\cite{Phillips:PV_EFT_2008}.)

While the contribution to the \SP scattering amplitude in Eq.~\eqref{AmpSP3} must be $\mu$-independent, this independence is achieved by the cancellation of the $\mu$-dependence  between the two terms in Eq.~\eqref{AmpSP3}.
Substituting Eq.~\eqref{CSP3} into Eq.~\eqref{AmpSP3} shows that: (i) The second term in Eq.~\eqref{CSP3} yields a $\mu$-independent contribution to  Eq.~\eqref{AmpSP3}; and (ii) the first term in Eq.~\eqref{CSP3} provides the $\mu$-dependence of the second term in  Eq.~\eqref{AmpSP3}  to cancel the $\mu$-dependence in the first term of Eq.~\eqref{AmpSP3}. 


Further, if there are no other as-yet-unidentified unnatural scales in the problem, $\bar{C}_3^{(SP)}$ should scale as $1/\nopicutoff^3$, again keeping in mind that this is the scaling with dimensionful parameters in the theory; the very different sizes of PV versus PC LECs is encoded in  a dimensionless proportionality constant involving $G_F m_\pi^2$, as discussed above.  The first term in Eq.~\eqref{CSP3} scales as $\frac{1}{M\Qmu^2 \nopicutoff^2}$ while the second term scales as $\frac{1}{M\Qmu \nopicutoff^3}$.  So long as $|\Qmu| = |1/a-\mu|  \ll  \nopicutoff$ it is useful to expand Eq.~\eqref{CSP3} as 
\beq
    C_3^{(SP)}(\mu) = \frac{\CSP_1(\mu)}{C_0(\mu)} C_{2}(\mu)\left[1 + O\left(\frac{\Qmu}{\nopicutoff}\right) \right].
    \label{CSP34}
\eeq
 In this case, the term with $\bar{C}_3^{(SP)}$ is suppressed by a factor of $\Qmu/\nopicutoff$ and can be neglected; $C_3^{(SP)}$  is dominated by the LO PV LEC $C_1^{(SP)}$ and the PC LECs $C_0$ and $C_2$. For $\Qmu\sim\nopicutoff$, which is the domain in which the large-\Nc relationships are expected to hold, both terms in Eq.~\eqref{CSP3} are expected to be of the same size,  and the expansion of Eq.~\eqref{CSP34} becomes irrelevant; $\bar{C}_3^{(SP)}$ cannot be neglected.


The \SP scattering amplitude up to and including three-derivative terms is given by 
\beq
\calA_{SP} = \calA^{[1]}_{SP} + \calA^{[3]}_{SP} \ ,
\label{eq:total_amplitude}
\eeq
where $\calA^{[1]}_{SP}$ and  $ \calA^{[3]}_{SP}$ correspond to the contributions from diagrams in Figs.~\ref{fig4} and \ref{three_derivative_diagrams_SP}, respectively, and are given by
\begin{align}
    \calA^{[1]}_{SP} & \propto -\frac{\CSP_1}{C_0} \frac{C_0p}{1-I_0C_0} = - \frac{4\pi }{M} \frac{p}{\LB \frac{1}{a}+ip \RB} \underbrace{\frac{\CSP_1}{C_0}}_{\text{LO}} \ ,       \label{eq:one_derivative_amp}
 \\
    \calA^{[3]}_{SP} &\propto -\frac{\CSP_1}{C_0}\frac{C_2p^3}{(1-I_0C_0)^2} - \bar{C}_3^{(SP)} \frac{C_0p^3}{1-I_0C_0} =
    - \frac{4\pi }{M} \frac{p}{\LB \frac{1}{a}+ip \RB} \left[  \underbrace{\frac{\CSP_1}{C_0}\frac{r_0 }{2} \frac{p^2}{\LB \frac{1}{a}+ip \RB}}_{\text{NLO}}  +\underbrace{\vphantom{\frac{\CSP_1}{C_0}\frac{r_0 }{2} \frac{p^2}{\LB \frac{1}{a}+ip \RB}} \bar{C}_3^{(SP)}p^2}_{\text{NNLO}} \right].
    \label{eq:three_derivative_amp}
\end{align}
The NLO correction to the PV scattering amplitude, which is suppressed by $Q/\nopicutoff$ relative to the LO contribution, is driven entirely by the leading PV LEC $\CSP_1$ and strong physics (the scattering length and the effective range parameter). 
No new undetermined PV parameter is required to this order; a new PV parameter is only required at NNLO. 


The PC \SP scattering amplitude up to and including NLO is given by \cite{Kaplan_KSW_1998a}

\begin{equation}
    \calA^{\text{PC}} = \mathcal{A}^\text{PC}_{-1} + \mathcal{A}^\text{PC}_{0} \propto \frac{4\pi}{M} \frac{1}{\LB \frac{1}{a}+ip \RB} \left[  1+ \frac{r_0}{2}\frac{p^2}{\LB\frac{1}{a}+ip\RB}  \right],
\end{equation}
where the subscripts $-1$ and $0$ indicate the scaling with $Q/\nopicutoff$ in the expansion. From Eqs.~\eqref{eq:one_derivative_amp}
and \eqref{eq:three_derivative_amp}, the PV scattering amplitude up to and including NLO is simply
\beq
\label{eq:PCPVrel}
\calA^{\text{PV}} \propto p \frac{\CSP_1}{C_0}\calA^{\text{PC}}.
\eeq
We can apply this result to the $\vec{p}p$ longitudinal asymmetry, given by \cite{Phillips:PV_EFT_2008}
\beq
A^{\vec{p}p}_L = \frac{\sigma_{+} - \sigma_{-}}{\sigma_{+} + \sigma_{-}} = 2\frac{\Re[\calA^{\text{PV}}\calA^{\text{PC}*}]}{|\calA^{\text{PC}}|^2} \ ,
\eeq
where for $\vec p p$ scattering, $ \calA^{\text{PV}}\sim C^{(\oneS-\Pzero)}_{1,~\Delta I=0}+C^{(\oneS-\Pzero)}_{1,~\Delta I=1}+C^{(\oneS-\Pzero)}_{1,~\Delta I=2} $, $\sigma$ is the total cross section and the subscript $+$ or $-$ indicates the beam helicity. From Eq.~\eqref{eq:PCPVrel} it is apparent that the LO asymmetry $A_L^{\vec{p}p}$ receives no NLO correction and only receives corrections at NNLO. This shows that the LO results of Ref.~\cite{Phillips:PV_EFT_2008} that fit a linear combination of single derivative LECs to the asymmetry $A_L^{\vec{p}p}$ hold through NLO. 

\subsection{\SP transitions at NNLO and beyond}

\begin{figure}[b]
    \centering
    \includegraphics[width=0.7\linewidth]{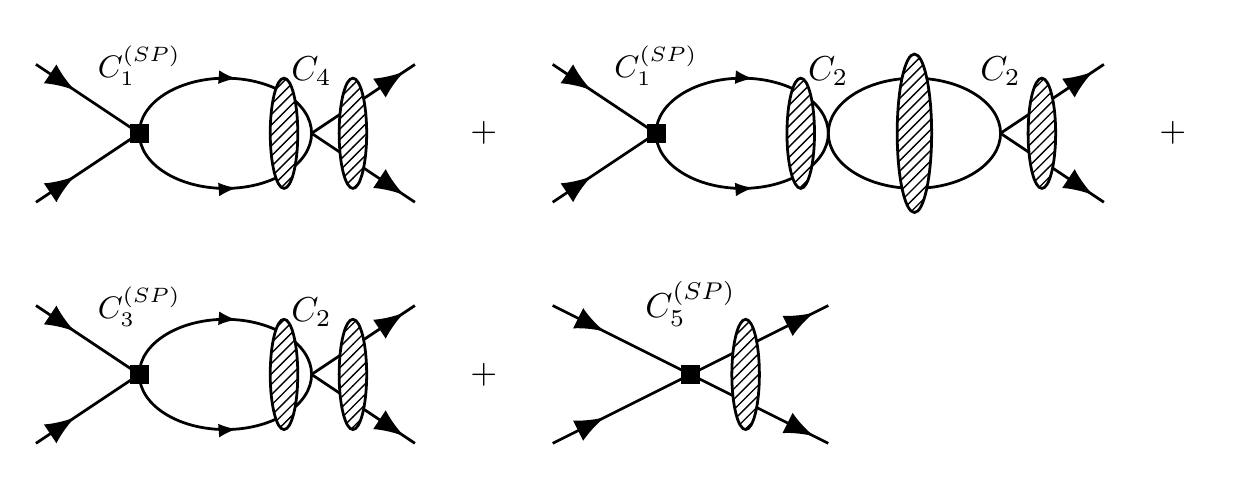}
    \caption{The  diagrams contributing to the five-derivative corrections to elastic \NN \SP transitions in \eftnopi. Notation as in Fig.~\ref{three_derivative_diagrams_SP}.
    \label{fig:5derivSP}}
\end{figure}

The total \SP scattering amplitude is given by
\beq
\calA_{SP} = \calA^{[1]}_{SP} + \calA^{[3]}_{SP} + \calA^{[5]}_{SP} + \cdots \ .
\eeq
 As shown in the previous subsection, there is a contribution to the NNLO \SP scattering amplitude that comes from $\bar{C}_3^{(SP)}$.  At the same order there will be contributions from  products of PV and PC operators containing a total of five derivatives. In particular, the scattering amplitude given by the sum of diagrams in Fig.~\ref{fig:5derivSP} yields, 
\beq
\calA^{[5]}_{SP} \propto -p^5 \LSB \CSP_1 \frac{I_0C_{4}}{(1-I_0C_0)^2} + \CSP_1 \frac{(I_0C_{2})^2}{(1-I_0C_0)^3} + \CSP_3 \frac{I_0C_2}{(1-I_0C_0)^2} + \CSP_5 \frac{1}{1-I_0C_0} \RSB,
\eeq
where  we have set $\CSP_5 \equiv C^{(S^{[0]}-P^{[5]})} + C^{(S^{[2]}-P^{[3]})} + C^{(S^{[4]}-P^{[1]})}$. Following the same procedure as above, we obtain the  RG equation 
\beq
\mu \dv{\mu}\CSP_5(\mu) = \frac{\mu M_N}{4\pi} \LSB \CSP_1(\mu) \  C_4(\mu)  + \CSP_3(\mu) \ C_2(\mu) + \CSP_5(\mu) \ C_0(\mu) \RSB.
\eeq
Solving this equation yields
\beq   
\CSP_5(\mu) = \frac{\CSP_1(\mu)}{C_0(\mu)}C_{4}(\mu) + \bar{C}^{(SP)}_3C_2(\mu) + \bar{C}^{(SP)}_5 C_0(\mu),
\eeq
where $\bar{C}^{(SP)}_5$ is a new constant of integration that, for naturally-sized boundary conditions, scales as $1/\nopicutoff^5$.  The resulting $p^5$ correction to the scattering amplitude is then given by
\begin{equation}
\begin{aligned}
    \calA^{[5]}_{SP}
    & \propto -p^5 \LSB \frac{\CSP_1}{C_0} \frac{(C_2)^2I_0}{(1-I_0C_0)^3} + \frac{\CSP_1}{C_0} \frac{C_4}{(1-I_0C_0)^2} + \bar{C}^{(SP)}_3\frac{C_2}{(1-I_0C_0)^2} + \bar{C}^{(SP)}_5\frac{C_0}{1-I_0C_0} \RSB
    \\
    & = -\frac{4\pi}{M}\frac{p^5}{\left( \frac{1}{a} + ip \right)} \LSB \underbrace{\frac{\CSP_1}{C_0} \frac{1}{\left( \frac{1}{a} + ip \right)^2}\frac{r_0^2}{4}}_{\text{NNLO}} + \underbrace{\vphantom{\frac{1}{\left( \frac{1}{a} + ip \right)^2}}\frac{\CSP_1}{C_0} \frac{1}{\left( \frac{1}{a} + ip \right)}\frac{r_1}{2} + \bar{C}^{(SP)}_3 \frac{1}{(\frac{1}{a}+ip)}\frac{r_0}{2}}_{\text{N$^3$LO}} + \underbrace{\vphantom{\frac{1}{\left( \frac{1}{a} + ip \right)^2}}\bar{C}^{(SP)}_5}_{\text{N$^4$LO}}\RSB .
\end{aligned}
\end{equation}
In the general case the RG equations of these PV LECs have the form\footnote{For a given partial-wave and isospin transition there exist several possible operator structures for a given power of momentum in the two-nucleon sector.  However, all of these operator structures reduce to a single operator structure for elastic \NN scattering.  For example, the operators associated with $C_4'$ and $\widetilde{C}_4$ both reduce to the operator for $C_4$.  The LECs in Eq.~\eqref{CSPRG} are for these single operator structures for each power of momentum in \NN scattering.}
\beq
\mu\dv{\mu} \CSP_{2n+1} = \frac{\mu M}{4\pi}\sum_{m=0}^n \CSP_{2(n-m)+1} C_{2m}.
\label{CSPRG}
\eeq
The complete solution to Eq.~\eqref{CSPRG} requires integration constants that must be fit to experiment or lattice calculations, which are currently unavailable. However, assuming that the integration constant $\Bar{C}^{(SP)}_{2n+1} \sim 1/\nopicutoff^{2n+1}$, the leading  behavior of PV LECs for  \murange is
\beqlist
\CSP_{2n+1}(\mu) & \equiv \sum_{m=0}^{n} C^{(S^{[2m]}-P^{[2n-2m+1]})}(\mu) \ \  \\
& = \frac{\CSP_1(\mu)}{C_0(\mu)} C_{2n}(\mu)\left[1+  O\left(\frac{\Qmu}{ \nopicutoff}\right) \right]
\label{eq35}.
\eeqlist
By analogy with the discussion of PC LECs in Ref.~\cite{Kaplan_KSW_1998a}, Eq.~\eqref{eq35} implies that the leading behavior of PV LECs for $n>0$ is driven only by the PV observable $\CSP_1/C_0$ and the $S$-wave strong interaction LECs. 
    But this does not mean that PV observables can be determined beyond NLO with $\CSP_1$ alone; at NNLO and beyond, higher-order PV parameters must be included.


\subsection{\DP transitions}
\begin{figure}[h]
    \centering
    \includegraphics[width=0.7\linewidth]{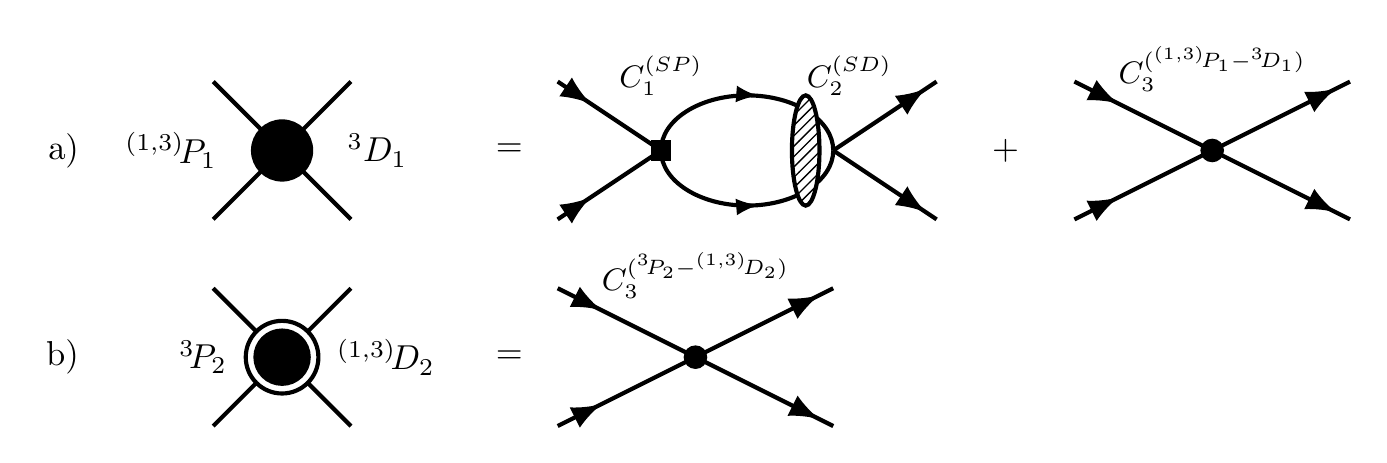}
    \caption{The  diagrams contributing to the \DP transitions in \eftnopi. The small squares denote insertions of PV operators with couplings $\CSP_1$, and the small circles denote insertions of PV operators with couplings $\CDP$ as indicated. The superscript (1,3) indicates either spin singlet or spin triplet. Diagrams with strong re-scattering on $P$-wave and $D$-wave channels occur at higher order.}
   \label{three_derivative_diagrams_PD}
\end{figure}

Now we consider the PV scattering amplitudes in the \DP channels. The four $J=2$ \DP channels only receive contributions from the tree diagram (Fig. \ref{three_derivative_diagrams_PD}b),
\beq
\calA^{[3]}_{\Ptwo-^{(1,3)}\!D_2} \propto  -p^3C^{(\Ptwo-^{(1,3)}\!D_2)}_3 \ ,
\label{eq: DP_amplitude}
\eeq
and  $\oneDthreePscalar$, $\oneDthreePvector$, $\oneDthreePtensor$, and $\threeDthreePtwo$ are each independent of $\mu$.
However, the $\oneP-\threeD$ ($\Delta I =0$) and $\threePone-\threeD$ ($\Delta I =1$) channels  receive contributions from loop diagrams with the {\SD}$\,$-mixing parameter (Fig. \ref{three_derivative_diagrams_PD}a). The corresponding scattering amplitudes are
\beq
    \calA^{[3]}_{^{(1,3)}\!P_1-\threeD}  \propto  -p^3C^{(^{(1,3)}\!P_1-\threeD)}_3 -p^3 C^{(\threeS-^{(1,3)}\!P_1)}_1 \frac{I_0\CSD}{1-I_0C_0} \ .
    \label{eq:AmpPD3}
\eeq
Requiring that $\mathcal{A}^{[3]}_{^{(1,3)}\!P_1-\threeD}$ be $\mu$-independent and using Eq.~\eqref{eq:CSD_beta_func} we obtain 
\beq
\mu\dv{\mu} C^{(^{(1,3)}\!P_1-\threeD)}_3(\mu) = \frac{\mu M}{4\pi} C^{(\threeS-^{(1,3)}\!P_1)}_1 (\mu)\CSD(\mu) \ ,
\eeq
yielding
\beq
C^{(^{(1,3)}\!P_1-\threeD)}_3(\mu) =  \frac{ C^{(\threeS-^{(1,3)}\!P_1)}_1(\mu)}{C_0^{(\threeS)}(\mu)}\CSD (\mu) + \Bar{C}^{(^{(1,3)}\!P_1-\threeD)}_3,
\label{eq:PDscaling}
\eeq
where $\Bar{C}^{(^{(1,3)}\!P_1-\threeD)}_3$  is another $\mu$-independent constant of integration. 
The structure is analogous to that of Eq.~\eqref{CSP3}.
Again, if the constant term is assumed to be of natural size and if \murangetriplet,   $\bar{C}^{(^{(1,3)}\!P_1-\threeD)}_3$ is suppressed by $\Qmu/\nopicutoff$, where $\Qmu = 1/a-\mu$,  compared to the other term in Eq.~\eqref{eq:PDscaling}.    
For this restriction on $\mu$, we can  write
\beq 
\label{eq:CDPrelCSP}
\frac{C^{(^{(1,3)}\!P_1-\threeD)}_3(\mu)}{\CSD(\mu)} \ \ =   \ \ \frac{ C^{(\threeS-^{(1,3)}\!P_1)}_1(\mu)}{C_0^{(\threeS)}(\mu)} \left[1+O\left(\frac{\Qmu}{\nopicutoff}\right)\right].
\eeq
This  implies that the leading contribution of $C^{(^{(1,3)}\!P_1-\threeD)}_3$  is  determined by PC LECs and a LO PV LEC.
A constraint on $C^{(\threeS-\threePone)}_1$ can be extracted from the recent NPDGamma measurement \cite{Blyth:2018aon}, using the result from Ref.~\cite{Schindler_PV_npdgamma_2010}. After adjusting for the different normalization of the operators, we find for \murangetriplet
\beqlist
    \frac{\threeDthreePone}{\CSD} + O\left(\frac{\Qmu}{\nopicutoff}\right)&\approx\frac{C^{(\threeS-\threePone)}_1}{C_0^{(\threeS)}}=(-7.3
    \pm 3.4 \pm 0.5 )\times 10^{-10} \textnormal{ MeV}^{-1} .
    \label{eq:PDconstraint}
\eeqlist
In the above equation, we list just the experimental (statistical plus systematic) errors from the NPDGamma experiment~\cite{Blyth:2018aon}. For the equality, the associated \eftnopi errors are $Q/\nopicutoff$.
Corrections to the approximation are roughly 30\% for $\Qmu\sim Q$ and include a contribution from $\bar{C}^{(\threePone-\threeD)}_3$ (they could be 100\% if $\Qmu$ is large).

The leading contributions to the $J=1$ \DP scattering amplitudes occur at order   $(Q/\nopicutoff)^2$ in the \eftnopi power counting. Substituting Eq.~\eqref{eq:PDscaling} into Eq.~\eqref{eq:AmpPD3}, they are given by
\begin{align}
    \calA^{[3]}_{^{(1,3)}\!P_1-\threeD}  
    & \propto   -\underbrace{\frac{4\pi}{M}\frac{p^3}{\frac{1}{a}+ip} \frac{C^{(\threeS-^{(1,3)}\!P_1)}_1(\mu) \ \CSD(\mu)}{\LSB C_0^{(\threeS)}(\mu)\RSB^2}} _{\text{N$^2$LO}}  - \underbrace{\vphantom{\frac{C^{(\threeS-^{(1,3)}\!P_1)}_1\CSD}{\LSB C_0^{(\threeS)}\RSB^2}} p^3\bar{C}^{(^{(1,3)}\!P_1-^3\!D_1)}_3}_{\text{N$^3$LO}}. 
    \label{eq:PD_amplitude}
\end{align}
In addition to the contribution from $\bar{C}^{(^{(1,3)}\!P_1-^3\!D_1)}_3$, at the same order there are contributions to the scattering amplitudes from products of PV and PC operators containing a total of five derivatives.

For processes involving the $\threeS$ channel it is useful to consider the effective range expansion about the deuteron pole \cite{PhysRev.76.38,PhysRev.77.647}. For this case, the PC LECs are expanded in powers of $Q/\nopicutoff$. For example, $C_0 = C_{0,-1}+C_{0,0}+ C_{0,1}\ldots$, where on the right hand side, the first and second subscripts indicate the number of derivatives of the operator, and the order  in powers of $Q/\nopicutoff$, respectively. The $\mu$-scaling of these PC LECs, including relativistic corrections, are given in Ref.~\cite{Chen:1999tn}. 
Similarly, we can  express the PV LECs as a power series in $Q/\nopicutoff$, e.g., $\CSP_1= \CSP_{1,-1}+\CSP_{1,0}+\CSP_{1,1}+\ldots$. We verified that the RG behaviors of the PV LECs in the  deuteron pole expansion can be analyzed order-by-order by following the same procedures we used above, and the arguments about which LECs are dominant in the range \murange are similar.

\section{Conclusions \label{sec6}}
We analyzed the three-derivative parity-violating \NN contact interactions in the dual \eftnopi and large-\Nc expansion. 
The minimal set of three-derivative \eftnopi operators consists of five that correct the (five) LO \SP transitions and six that describe the leading contribution to the  \DP transitions. 
The LECs accompanying these operators are free parameters in \eftnopi  and must be fit to experimental data or extracted from lattice QCD. 
The large-\Nc expansion reduces the number of independent LECs at a given order either by demoting some to higher order in the dual \eftnopi and large-\Nc expansion, or by relating some LECs to each other. 
By organizing the LECs of the three-derivative operators in \eftnopi in powers of 1/\Nc, we can establish a hierarchy among them. The  isoscalar and  isotensor LECs appear at LO in \Nc, while  four isovector LECs appear at subleading order in \Nc. 
The relationships found among the LECs hold up to corrections of $1/\Nc^2\sim 10\%$ channel-by-channel. The relationships among the LECs for three-derivative \SP operators  echo those found in the one-derivative sector; in both cases, the two isoscalar LECs  are not independent and isoscalar and isotensor terms dominate in the large-\Nc limit. 
For elastic scattering, the eleven independent LECs that occur at three derivatives  in \eftnopi are reduced to two at LO in the dual \eftnopi and large-\Nc expansion. This may provide constraints that can be checked by future experiments and lattice QCD calculations. As usual, because $\Nc=3$ in our world, a large-\Nc treatment cannot offer exact predictions, but  should indicate general trends. 

We also studied the RG behavior of the three-derivative operators and analyzed the running of the  PV LECs with respect to the subtraction point $\mu$. For $\mu$ in the range \murange, the leading behavior of all the \SP LECs only depends on  $\CSP_1$ and the strong $S$-wave LECs. 
The analysis of the \DP LECs shows that $\threeS-\threeD$ mixing introduces $\mu$ dependence in the $J=1$ \DP LECs, while the $J=2$ \DP LECs are $\mu$-independent. 
This RG analysis allows us to power count PV \NN scattering operators in  \eftnopi; it indicates that some previous (LO) calculations of \SP channel observables \cite{Phillips:PV_EFT_2008} do not  obtain corrections until NNLO.  It also shows that LECs in two $J=1$ \DP channels are related to  \SP LECs for sufficiently small values of $\mu$. 
We emphasize again that the value of $\mu$ needed for large-\Nc relationships to hold is incompatible with the small values of $|1/a-\mu|$ that lead to some simplifications of the RG relationships.

\begin{acknowledgments}

The request to find large-\Nc constraints on \DP wave operators was made by the participants of the KITP Workshop on Hadronic Parity Nonconservation (2018) \cite{KITP_2018}. STN thanks H.~Singh for many insightful suggestions and guidance throughout this project. RPS thanks H. Singh for useful discussions and for his mentorship of STN.   This material is based upon work supported by the U.S. Department of Energy, Office of Science, Office of Nuclear Physics, under Award Number DE-FG02-05ER41368 (STN and RPS), and Award Number DE-SC0019647 (MRS). 
\end{acknowledgments}

\appendix

\section{Fierz transformation \label{app2}}
In this section, we show in detail one method for matching between the large-\Nc and partial-wave bases using Fierz identities. The relevant Fierz identities are  
\beqlist
    \delta_{AB}\delta_{CD} = & ~ \frac{1}{2}(\sigma_2)_{CA} (\sigma_2)_{BD} + \frac{1}{2}(\sigma_i\sigma_2)_{CA} (\sigma_2\sigma_i)_{BD},\\
    (\sigma_i)_{AB} \delta_{CD} = & - \frac{1}{2} \LSB(\sigma_i\sigma_2)_{CA}(\sigma_2)_{BD}+(\sigma_2)_{CA} (\sigma_2\sigma_i)_{BD}\RSB - \frac{1}{2} i\epsilon_{ijk} (\sigma_j\sigma_2)_{CA} (\sigma_2\sigma_k)_{BD},\\
    (\sigma_i)_{AB}(\sigma_j)_{CD}=& -\frac{1}{2}\delta_{ij}(\sigma_2)_{CA}(\sigma_2)_{BD}+\frac{1}{2}i\epsilon_{ijk}\big[(\sigma_k\sigma_2)_{CA}(\sigma_2)_{BD} -(\sigma_2)_{CA}(\sigma_2\sigma_k)_{BD}\big]\\
    &-\frac{1}{2}(\delta_{ik}\delta_{jn}+\delta_{in}\delta_{jk}-\delta_{ij}\delta_{kn})(\sigma_k\sigma_2)_{CA}(\sigma_2\sigma_n)_{BD},
\label{fierz}
\eeqlist
where the upper case Latin letters denote spin indices. These formulas are also applied to isospin matrices by  substituting $\sigma\rightarrow\tau$.

\begin{figure}[ht!]
    \centering
    \includegraphics[width=0.3\linewidth]{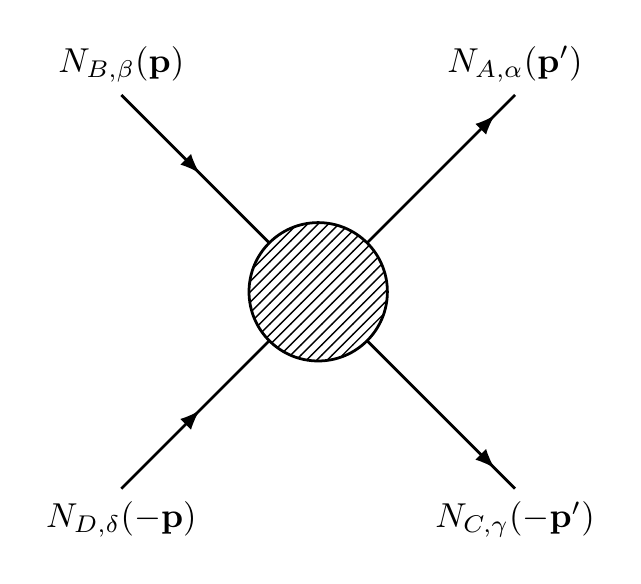}
    \caption{Assignment of spin, isospin indices and momenta.  Greek letters denote isospin indices and capital Latin letters denote spin indices. \label{fig1} }
\end{figure}
One approach to performing  Fierz transformations is to take matrix elements, as outlined in Appendix A of Ref.~\cite{Phillips:PV_EFT_2008}. For example, the matrix element of the operator corresponding to $\widetilde{C}^{[3]}_{G.G}$ in   Eq.~\eqref{NLO_weinberg_Lag} (see Fig.~\ref{fig1} for momentum and index assignments) is
\beqlist
\langle\widetilde{\calO}^{[3]}_{G.G}\rangle & =  2\epsilon_{ijk}[(\sigma_i)_{AB}(\sigma_l)_{CD} + (\sigma_l)_{AB}(\sigma_i)_{CD}]\isospinvector (\tau_a)_{\alpha \beta}(\tau_b)_{\gamma \delta}(p_i + p_i')(p_k - p'_k)(p_l - p'_l)\\
&-2\epsilon_{ijk}[(\sigma_i)_{CB}(\sigma_l)_{AD} + (\sigma_l)_{CB}(\sigma_i)_{AD}]\isospinvector (\tau_a)_{\gamma \beta}(\tau_b)_{\alpha \delta}(p_i - p_i')(p_k + p'_k)(p_l + p'_l).
\eeqlist
The operator symbol in the matrix element above echoes its corresponding LEC; e.g., $\widetilde{\calO}^{[3]}_{G.G}$ is the operator associated with $\widetilde{C}^{[3]}_{G.G}$. Applying the identities in Eq.~\eqref{fierz}  gives
\beq
\langle\widetilde{\calO}^{[3]}_{G.G}\rangle =  64 ~ i\epsilon_{ijk} (\delta_{im}\delta_{ln} + \delta_{in}\delta_{lm}) (P^\dagger_{m})_{CA,\gamma\alpha} (P_{n,3})_{BD,\beta\delta} (p_jp'_kp'_l) + \hc.
\label{eqB5}
\eeq
This structure  suggests that the $\widetilde{C}^{[3]}_{G.G}$ term   may contribute  to  $\threeS-\threePone$, $\threePone-\threeD$, and $\Ptwo-\threeDtwo$  transitions. The corresponding partial-wave matrix elements for these three channels are
\begin{align}
\langle\calO^{[3]}_{\threeS-\threePone}\rangle = & -4\threeSthreeP ~i \epsilon_{ijk} (P^\dagger_i)_{CA,\gamma\alpha} (P_{j,3})_{BD,\beta\delta} (p'^2p_k) + \hc,\\
\langle\calO^{[3]}_{\threePone-\threeD}\rangle = & -4\threeDthreePone ~i \epsilon_{ijk} (P^\dagger_l)_{CA,\gamma\alpha} (P_{i,3})_{BD,\beta\delta} (p'_lp'_kp_j) \\\nonumber
& + \frac{4}{3}\threeDthreePone ~i \epsilon_{ijk} (P^\dagger_i)_{CA,\gamma\alpha} (P_{j,3})_{BD,\beta\delta} (p'^2p_k) + \hc,\\
\langle\calO^{[3]}_{\Ptwo-\threeDtwo}\rangle = & -4\threeDthreePtwo ~i \epsilon_{ijk} (P^\dagger_i)_{CA,\gamma\alpha} (P_{j,3})_{BD,\beta\delta} (p'_kp'_lp_l)  \\\nonumber
& - 4\threeDthreePtwo ~i \epsilon_{ijk} (P^\dagger_i)_{CA,\gamma\alpha} (P_{l,3})_{BD,\beta\delta} (p'_kp'_lp_j) + \hc,
\end{align}
where the projectors $P_x$ are defined in Eq.~\eqref{proj}. Using  
\beqlist
\epsilon_{ijk}(P^\dagger_i)_{CA,\gamma\alpha}(P_{j,3})_{BD,\beta\delta}(p'_kp'_lp_l) & = \epsilon_{ijk}(P^\dagger_i)_{CA,\gamma\alpha}(P_{j,3})_{BD,\beta\delta}(p_kp'^2) \\
&+ \epsilon_{ijk}(P^\dagger_i)_{CA,\gamma\alpha}(P_{l,3})_{BD,\beta\delta}(p'_kp'_lp_j) \\
& - \epsilon_{ijk}(P^\dagger_l)_{CA,\gamma\alpha}(P_{i,3})_{BD,\beta\delta}(p'_kp'_lp_l)
\label{eqB9}
\eeqlist
with Eq. (\ref{eqB5})  yields
\beq
\langle\widetilde{\calO}^{[3]}_{G.G}\rangle = - 24\langle\calO^{[3]}_{\threePone-\threeD}\rangle -8\langle\calO^{[3]}_{\Ptwo-\threeDtwo}\rangle .
\eeq

\bibliography{ref.bib}
\end{document}